Bi-annual Journal of Gwalior Academy of Mathematical Sciences
Print ISSN 0974-2689* Online ISSN 1945-9181
Vol. 6(a) * Nos.1 * December 2018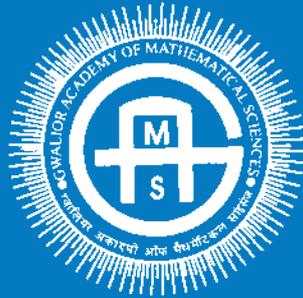

# GAMS Journal of Mathematics and Mathematical Biosciences

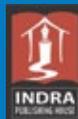

Indra Publishing House
www.indrapublishing.com

# Sensitivity Analysis of SEIR – SEI Model of Dengue Disease

Ganga Ram Phaijoo[1] and Dil Bahadur Gurung[2]
[1,2] Department of Natural Sciences (Mathematics), School of Science
Kathmandu University, Kavre, Nepal
*Email: gangaram@ku.edu.np

**Abstract:** Dengue is a vector borne infectious disease. The disease is transmitted by *Aedes* mosquitoes. In the present work, SEIR - SEI compartmental epidemiological model is used to describe dengue disease transmission dynamics. The human population is classified into four epidemiological states: Susceptible, Exposed, Infectious and Recovered humans. The mosquito population is divided into three epidemiological states: Susceptible, Exposed and Infectious mosquitoes. Associated basic reproduction number which determines whether the disease will persist or die out, is calculated by Next Generation method. Stability of the equilibrium points of the model is analyzed in terms of basic reproduction number. Sensitivity analysis is carried out to study importance of model parameters.

## 1. INTRODUCTION

Infectious diseases have been spreading all over the world. Mathematical models are influential tools to understand transmission dynamics of the infectious disease, to predict the future outbreak of the disease and to propose control strategies of the disease. So, many mathematical studies have been made to study infectious diseases. Kermack and Mckendrick in 1927, proposed an SIR compartmental model to study transmission dynamics of infectious diseases [1].

Dengue is a vector borne infectious disease that is affecting humans of almost all ages worldwide. The disease is transmitted to humans by the bites infected *Aedes* mosquitoes called *Aedes aegypty* and *Aedes albopictus*. Four serologically different but closely related viruses DEN 1 to DEN 4 cause dengue disease [2, 3]. Esteva and Vargas used SIR model to study dengue disease mathematically taking constant [4] and variable [5] human population. Nuraini et al. took Severe DHF compartment into account to study transmission dynamics of dengue disease [6]. Pinho et al. made a comparative study of two dengue epidemic cases that occured in Salvador, Brazil in 1995 - 1996 and 2002 using real data [7]. Phaijoo and Gurung, Gakkhar and Chavda used SIR model to analyse the impact of awareness in transmission dynamics of dengue disease [8, 9]. Sardar et al. introduced memory in the compartmental model of dengue disease transmission between human and mosquito. They incorporated memory in the model by the use of fractional differential operator [10].

Both extrinsic and intrinsic incubation periods play significant role in the dynamics of dengue disease. So, many mathematical studies of dengue disease have been made using the compartment of incubation period. Side and Noorani proposed SEIR model of dengue disease taking exposed classes of both vector and host population [11]. Pongsumpun analysed transmission dynamics of dengue disease with and without effect of extrinsic incubation period [12]. Chan and Johansson studied the extrinsic and intrinsic incubation periods in mosquito and human respectively [13]. Phaijoo and Gurung studied the impact of temperature and human movement on the persistence of dengue disease using SEIR model in multi-patch environment [14].

While studying the infectious diseases mathematically, it is always required to determine important model parameters influencing the transmission of the disease. So, sensitivity analysis is performed in the mathematical studies. Chitnis et al. determined important parameters in the spread of malaria through sensitivity analysis mathematically [15]. Shah and Gupta, Rodrigues et al. determined relative importance of the model parameters to the disease transmission dynamics [16].

## 2. FORMULATION AND DESCRIPTION OF THE MODEL

For the formulation of the model, we divide the total host population $N_h$ to four classes: $S_h$ (Susceptible), $E_h$ (Exposed), $I_h$ (Infectious) and $R_h$ (Recovered); and the total vector population $N_v$ to three classes: $S_v$ (Susceptible), $E_v$ (Exposed) and $I_v$ (Infectious). The recovered class in vector population is not considered because infection period in mosquitoes end with their death.

The birth rate for the host population is $\mu_h$ (assumed to be equal to the death rate). The susceptible hosts get infected (exposed state) by the bite of infectious vectors at the rate of $\frac{b\beta_h I_v}{N_h}$ where $b$ is the biting rate of the vector, $\beta_h$ is the transmission probability of dengue from vectors to the hosts. The exposed hosts either die due to the natural cause or move the infectious class at the rate $v_h$ after showing the clinical symptoms of dengue disease. Infectious hosts either die due the natural cause or recover at the rate $\gamma_h$.

Similarly, susceptible vectors move to the exposed class at the rate of $\frac{b\beta_v I_h}{N_h}$ and they move to the exposed class at the rate of $v_v$ after showing the clinical symptoms of the disease. Mosquitoes are recruited at the rate of $\pi_v$, and $\mu_v$ is the death rate of mosquitoes.

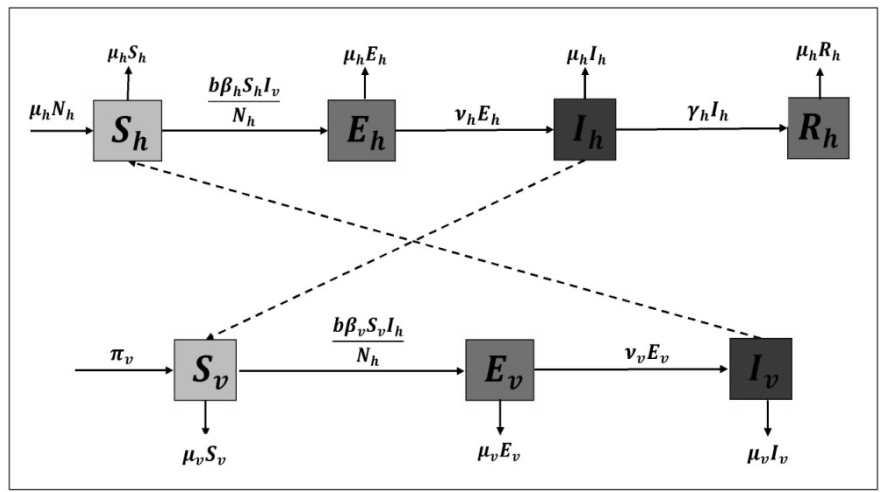

**Figure 1:** *Flow chart of the SEIR - SEI dengue epidemiological model.*

Figure 1 describes the SEIR – SEI model of dengue disease. The system of differential equations which describe the model is

$$\frac{dS_h}{dt} = \mu_h N_h - \frac{b\beta_h}{N_h} S_h I_v - \mu_h S_h$$

$$\frac{dE_h}{dt} = \frac{b\beta_h}{N_h} S_h I_v - (\nu_h + \mu_h) E_h$$

$$\frac{dI_h}{dt} = \nu_h E_h - (\gamma_h + \mu_h) I_h$$

$$\frac{dR_h}{dt} = \gamma_h I_h - \mu_h R_h \tag{2.1}$$

$$\frac{dS_v}{dt} = \pi_v - \frac{b\beta_v}{N_h} S_v I_h - \mu_v S_v$$

$$\frac{dE_v}{dt} = \frac{b\beta_v}{N_h} S_v I_h - (\nu_v + \mu_v) E_v$$

$$\frac{dI_v}{dt} = \nu_v E_v - \mu_v I_v$$

Here,

Total hostpopulation,

$$N_h = S_h + E_h + I_h + R_h,$$

Total vector population

$$N_v = S_v + E_v + I_v.$$

Also,

$$\frac{dN_h}{dt} = 0, \quad \frac{dN_v}{dt} = \pi_v - \mu_v N_v$$

So, $N_h$ remains constant and $N_v$ approaches $\frac{\pi_v}{\mu_v}$ when $t \to \infty$.

Introducing the proportions:

$$s_h = \frac{S_h}{N_h}, \; e_h = \frac{E_h}{N_h}, \; i_h = \frac{I_h}{N_h}, \; r_h = \frac{R_h}{N_h}, \; s_v = \frac{S_v}{A/\mu_v}, \; e_v = \frac{E_v}{A/\mu_v}, \; i_v = \frac{I_v}{A/\mu_v}$$

Here,

$$r_h = 1 - s_h - e_h - i_h \text{ and } s_v = 1 - e_v - i_v.$$

So, the system of equations (2.1) has the same qualitative behavior as the following system of ordinary differential equations

$$\frac{ds_h}{dt} = \mu_h(1 - s_h) - \alpha s_h i_v$$

$$\frac{de_h}{dt} = \alpha s_h i_v - \beta e_h$$
$$\frac{di_h}{dt} = \nu_h e_h - \gamma i_h \qquad (2.2)$$
$$\frac{de_v}{dt} = \delta s_v i_h - (\epsilon + \nu_v) e_v$$
$$\frac{di_v}{dt} = \nu_v e_v - \epsilon i_v$$

Here,

$$\alpha = \frac{b\beta_h \pi_v}{N_h \mu_v}, \qquad \beta = \nu_h + \mu_h, \qquad \gamma = \gamma_h + \mu_h, \qquad \delta = b\beta_v, \qquad \epsilon = \mu_v$$

## 3. EQUILIBRIUM POINTS AND STABILITY ANALYSIS

Equilibrium Points: There are two equilibrium points of the system of equations (2.2) namely
$$E_0 = (1, 0, 0, 0, 0) \text{ and } E_1 = (s_h^*, e_h^*, i_h^*, e_v^*, i_v^*)$$
where,

$$s_h^* = \frac{(\beta\gamma\epsilon + \delta\mu_h\nu_h)(\epsilon + \nu_v)}{\delta\nu_h[\epsilon\mu_h + (\alpha + \mu_h)\nu_v]]}, \quad e_h^* = \frac{\mu_h\gamma\epsilon(\epsilon + \nu_v)(R_0^2 - 1)}{\delta\nu_h[\epsilon\mu_h + (\alpha + \mu_h)\nu_v]},$$

$$i_h^* = \frac{\mu_h\epsilon(\epsilon + \nu_v)(R_0^2 - 1)}{\delta[\epsilon\mu_h + (\alpha + \mu_h)\nu_v]},$$

$$e_v^* = \frac{\mu_h\beta\gamma\epsilon^2(R_0^2 - 1)}{\delta(\beta\gamma\epsilon + \delta\mu_h\nu_h)}, \qquad i_v^* = \frac{\mu_h\beta\gamma\epsilon(R_0^2 - 1)}{\delta(\beta\gamma\epsilon + \delta\mu_h\nu_h)}$$

Also, $R_0$ is basic reproduction number defined as the spectral radius of the matrix $FV^{-1}$ i.e. $\rho(FV^{-1})$ ($F$ is the matrix of transmission terms and $V$ is the matrix of transition terms). The expression for $R_0$ computed using Next Generation matrix method [17, 18] is obtained as

$$R_0 = \sqrt{\frac{\alpha\delta\nu_h\nu_v}{\beta\gamma\epsilon(\epsilon + \nu_v)}}$$

Here,

$$F = \begin{bmatrix} 0 & 0 & 0 & \alpha \\ 0 & 0 & \delta & 0 \\ 0 & 0 & 0 & 0 \\ 0 & 0 & 0 & 0 \end{bmatrix} \text{ and } V = \begin{bmatrix} \beta & 0 & 0 & 0 \\ 0 & \epsilon + \nu_v & 0 & 0 \\ -\nu_h & 0 & \gamma & 0 \\ 0 & -\nu_v & 0 & \epsilon \end{bmatrix}$$

The first equilibrium point $E_0$ which always exists is called disease free equilibrium (DFE) and the second point $E_1$ if exists is called endemic equilibrium point. The second point exists if $R_0 > 1$.

We have the following results:

**Theorem 3.1** (Existence of Equilibrium Points). *The System of equations (2.2) always has a disease free equilibrium point. If $R_0 > 1$, the system has a unique endemic equilibrium point.*

**Theorem 3.2** (Local stability of DFE). *Disease free equilibrium of the system of equations (2.2) is locally asymptotically stable if $R_0 < 1$ and unstable if $R_0 > 1$.*

*Proof:* Jacobian matrix of the system of equations (2.2) at disease free equilibrium is obtained as the block structure

$$J = \begin{bmatrix} J_{11} & J_{12} \\ 0 & F - V \end{bmatrix}$$

Here, eigenvalue of the matrix $J_{11} = [-\mu_h]$ is $-\mu_h < 0$. So, the stability of the DFE depends on the matrix $F - V$ where the matrix $F$ is non-negative and $V$ is non-singular $M$ – matrix [19]. Again, spectral abscissa of the matrix $F - V$, $s(F - V) < 0 \Leftrightarrow \rho(FV^{-1}) = R_0 < 1$ [18]. So, the DFE is locally asymptotically stable if $R_0 < 1$.

$R_0 > 1 \Rightarrow s(F - V) > 0$. It implies that at least one eigenvalue will have positive real part. So, the DFE is unstable if $R_0 > 1$.

**Theorem 3.3** (Global stability of DFE). *Disease free equilibrium of the system of equations (2.2) is globally asymptotically stable if $R_0 < 1$.*

*Proof:* In the present model, $s_h \leq 1$, $s_v \leq 1$. So, from the system of equations (2.2), for the dynamics of infective population

$$\begin{aligned} \frac{de_h}{dt} &\leq \alpha i_v - \beta e_h \\ \frac{de_v}{dt} &\leq \delta i_h - (\epsilon + \nu_v) e_v \\ \frac{di_h}{dt} &= \nu_h e_h - \gamma i_h \\ \frac{di_v}{dt} &= \nu_v e_v - \epsilon i_v \end{aligned} \quad (3.1)$$

The corresponding linear system of equations is

$$\begin{aligned} \frac{de_h}{dt} &= \alpha i_v - \beta e_h \\ \frac{de_v}{dt} &= \delta i_h - (\epsilon + \nu_v) e_v \\ \frac{di_h}{dt} &= \nu_h e_h - \gamma i_h \\ \frac{di_v}{dt} &= \nu_v e_v - \epsilon i_v \end{aligned} \quad (3.2)$$

The system of linear equations (3.2) can be written as

$$\frac{d\vec{u}}{dt} = K \vec{u} \quad (3.3)$$

where $K = F - V$ and $\vec{u} = [e_h, e_v, i_h, i_v]^T$.

Since $F$ is a non - negative matrix and $V$ is a non - singular $M$ - matrix, if $R_0 = \rho(FV^{-1}) < 1$, then $s(F - V) < 0$[18]. So, each positive solution of (3.2) satisfies

$\lim_{t\to\infty} e_h = 0, \lim_{t\to\infty} e_v = 0, \lim_{t\to\infty} i_h = 0,$ and $\lim_{t\to\infty} i_v = 0$.

Hence, DFE of the system of equations (3.2) is globally asymptotically stable since the system is linear. Since all the variables in the system of equations (2.2) are nonnegative, the use of a comparison theorem [20, 21] leads

$\lim_{t\to\infty} e_h = 0, \lim_{t\to\infty} e_v = 0, \lim_{t\to\infty} i_h = 0, \lim_{t\to\infty} i_v = 0$ and $\lim_{t\to\infty} s_h = 1$.

Hence, the DFE, $(1, 0, 0, 0, 0)$ is globally asymptotically stable if $R_0 < 1$.

## 3. SENSITIVITY ANALYSIS

We use normalized sensitivity index [15] for sensitivity analysis. The normalized forward sensitivity index of variable $\xi$ which depends on a parameter $\zeta$ is defined as

$$\Upsilon_\xi^\zeta = \frac{\partial \zeta}{\partial \xi} \times \frac{\xi}{\zeta}$$

We take $\zeta = R_0$ and $\xi = \pi_v, b, \beta_v, \mu_v, \nu_v, \beta_h, \gamma_h, \mu_h, \nu_h$.

Table 1: Sensitivity analysis of model parameters

| Parameters | Baseline Values | Sensitivity Indices |
|---|---|---|
| $\pi_v$ | 5000 | + 0.5 |
| $b$ | 0.5 | + 1 |
| $\beta_v$ | 1 | + 0.5 |
| $\mu_v$ | 0.02941 | - 1.31823 |
| $\nu_v$ | 0.1428 | + 0.318228 |
| $\beta_h$ | 0.75 | + 0.5 |
| $\gamma_h$ | 0.328833 | - 0.49993 |
| $\mu_h$ | 0.0045 | - 0.000208 |
| $\nu_h$ | 1.667 | + 0.000138 |

Table 1 shows the sensitivity index values of the model parameters. Positive values describe that the prevalence of the disease increases with the increase in the parameter values. Sensitivity indices of the parameters $\nu_v$, $\gamma_h$ and $\mu_h$ are negative. So, they contribute in decreasing the value of basic reproduction number and so, contribute in decreasing the prevalence of the disease. Also, the Tableshows that the most positive sensitive parameter is the biting rate of the mosquito and the most negative sensitive model parameter is death rate of the mosquito.

## 4. NUMERICAL RESULTS AND DISCUSSION

Simulations are carried out in order to observe the effects of model parameters on the transmission dynamics and spread of dengue disease. Following numerical values are used for the simulations:

$N_h = 5071126$, $\nu_h = 0.1667$, $\mu_h = 0.000046$, $\pi_v = 2500000$, $\beta_h = 0.75$, $\gamma_h = 0.328833$, $\beta_v = 0.375$, $\nu_v = 0.1428$, $\mu_v = 0.25$, $b = 0.5$ [11].

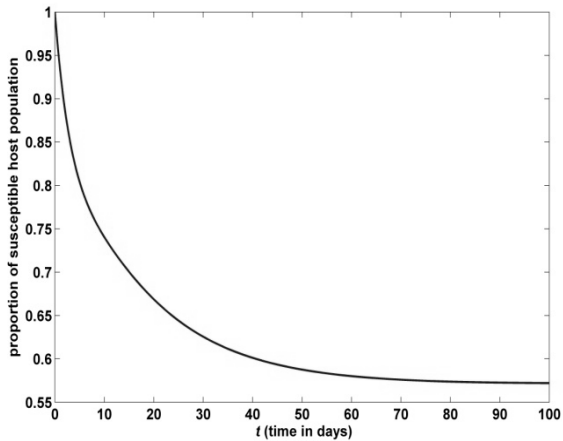

Figure 1: *Dynamics of susceptible host population.*

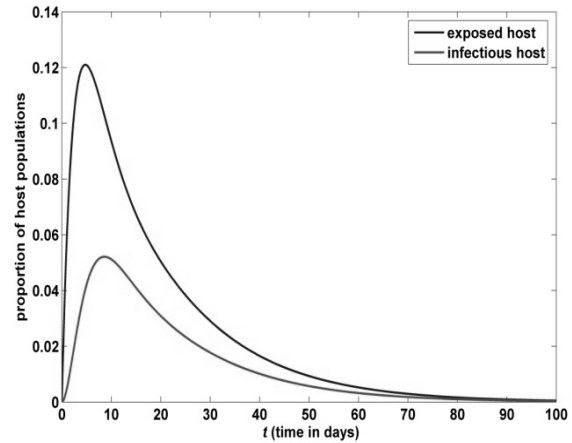

Figure 2: *Dynamics of infected host populations.*

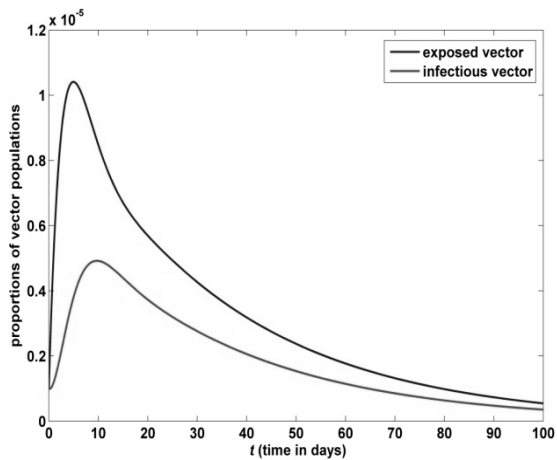

Figure 3: *Dynamics of infected vector populations.*

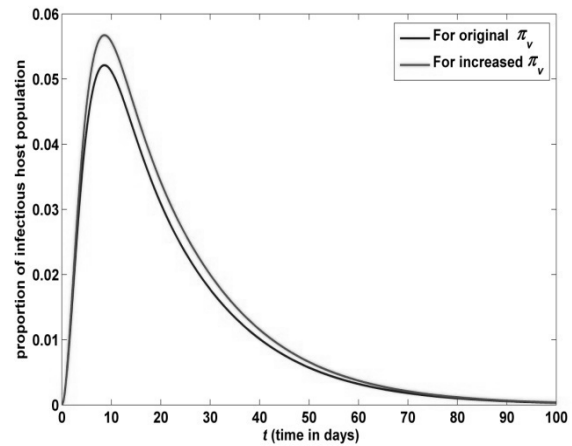

Figure 4: *Effect of $\pi_v$ on $I_h$.*

Figure 1 to Figure 3 is simulated to investigate the transmission dynamics of dengue disease. It is observed that susceptible host population decreases over the time. It is because susceptible hosts get infected due to the bite of infectious mosquitoes and some of them die due to the natural cause (Figure 1). The infected (exposed) host population increases initially due to the interaction of susceptible host with infectious vectors. Later, the exposed hosts move to the infectious class showing the symptoms of the disease. So, the population starts decreasing. The infectious host population decreases due to the recovery from the disease and due to death from natural cause.

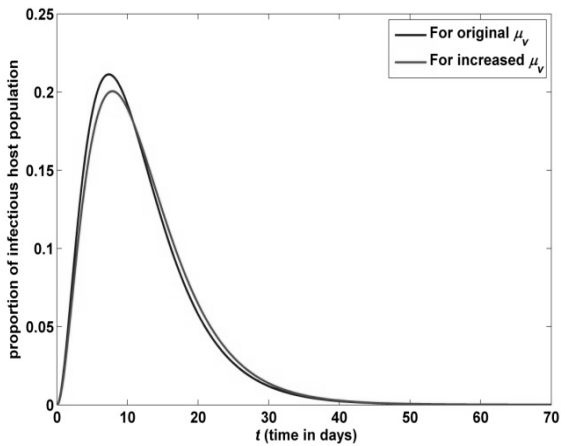
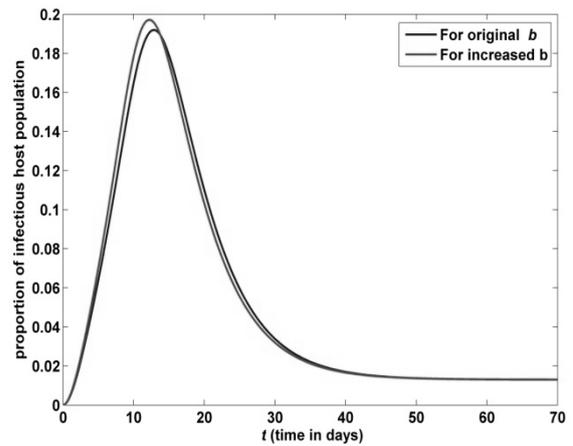

Figure 5: *Effect of $\mu_v$ on $I_h$.*   Figure 6: *Effect of b on $I_h$.*

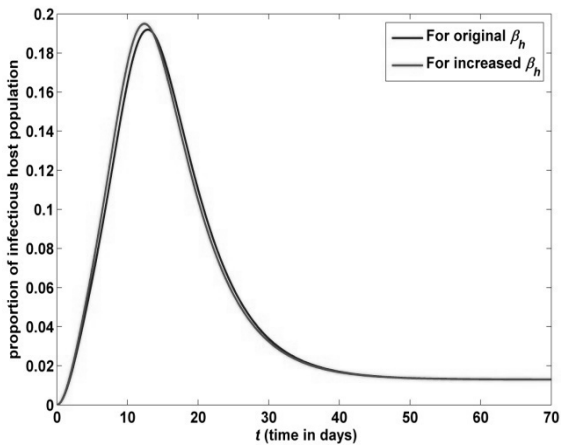
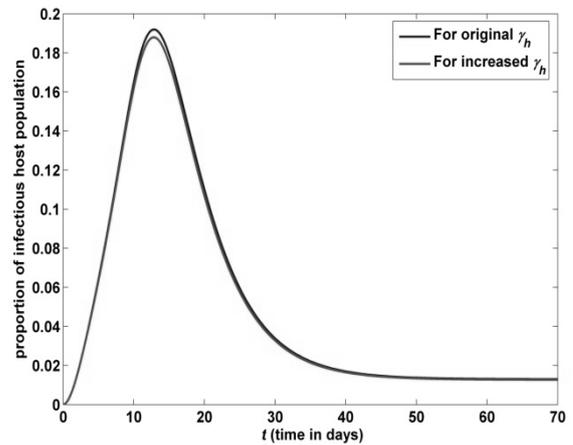

Figure 7: *Effect of $\beta_h$ on $I_h$.*   Figure 8: *Effect of $\gamma_h$ on $I_h$.*

We have simulated Figure 4 to Figure 8 to investigate the sensitivity of model parameters. We see that the population size of infectious hosts is increasing with the increased values of model parameters $\pi_v$, $b$, $\beta_h$. These parameters have positive sensitivity indices (Table 1). Also, the parameters $\mu_v$ and $\gamma_h$ are contributing in decreasing the number of infectious hosts. These parameters have negative sensitivity indices. Thus, it is observed that parameters with positive sensitivity index increases the disease transmission and the parameters with negative sensitivity index decreases the transmission of the disease. The simulated results and the Table 1 show that $\mu_v, b, \gamma_h$ are more sensitive model parameters. $b$ is the most positive sensitive parameter which increases the transmission of the disease and $\mu_v$ is the most negative sensitive parameter which decreases the transmission of the disease.

## 4. NUMERICAL RESULTS AND DISCUSSION

In the present work, we have discussed SEIR - SEI model of dengue disease taking exposed class in both host and vector populations. We performed sensitivity analysis to determine the important model parameters which affect the transmission dynamics of dengue disease significantly. We observed that the most important sensitive parameters are biting rate $b$ (positive) and death rate $\mu_v$ (negative) of the mosquitoes. Increasing biting rate increases the transmission of the disease and increasing death rate decreases the transmission of the disease significantly. So, by increasing death rate and decreasing the biting rate of the mosquitoes we can decrease the prevalence of dengue disease.